# Time Eigenvalues For The One-dimensional Infinite Square Well
David M. Rosenbaum

*Discrete time eigenvalues exist for the one-dimensional infinite square well. This paper finds the values and describes the associated eigenfunctions in detail.*

INTRODUCTION

This paper, and ultimately all quantum mechanics, is based on the three dimensional commutation relation

$$[p,q] = -i\hbar. \qquad (1)$$

The fourth component of the commutation relation would naturally be

$$[H,T] = -i\hbar, \qquad (2)$$

where H is the Hamiltonian and T is a Hermitian time operator.

As Pauli[1] pointed out in the 1920s, no such time operator can exist in Hilbert space unless H has a continuum of eigenvalues from $-\infty$ to $+\infty$.

However, for a long time, quantum mechanics has not really been done in Hilbert Space. For example, $e^{ikx}$, $|x|$ and δ(x) do not exist in $L_2$.

In 1969 I published a paper[2] in which physical states are represented by continuous linear functionals on a space of good functions[3], rather than by functions in a Hilbert space. Since $L_2$ is isomorphic to a subset of Super Hilbert space, everything that can be done in $L_2$ can be done in Super Hilbert space. In addition, lots of other things exist in Super Hilbert space, such as delta-functionals and time operators.

This paper is about the one-dimensional infinite square well, so the calculations are in one dimension. The extensions to two and three dimensions are straightforward.

---

[1]. W. Pauli, <u>Handbuch der Physik</u>, Vol. 24/1, page 143

[2]. David M. Rosenbaum, Super Hilbert Space and the Quantum-Mechanical Time Operators, J. Math. Phys. 1127 (1969)

[3] Good functions are functions which are everywhere differentiable any number of times and such that they and their derivatives fall off at infinity faster than the inverse of any polynomial.

MOMENTUM REPRESENTATIONS

No representations of operators were used in reference (2), so it is important to show that the results in this paper are independent of the representation chosen.

The representation of p is restricted only by (1), so let

$$p \to -i\hbar \frac{d}{dx} + f(x) \;, \tag{3}$$

where the arrow stands for "be represented by" and $f(x)$ is any real function with a first derivative and an indefinite integral.

Energy

With this representation of p, the energy eigenvalue equation for a potential V(x) is:

$$\frac{d^2\psi}{dx^2} + i\left(\frac{2}{\hbar}\right)f(x)\frac{d\psi}{dx} + \frac{1}{\hbar^2}\left[i\hbar\frac{df}{dx} - f^2(x) - 2mV(x) + 2mE\right]\psi(x) = 0 \;. \tag{4}$$

Let

$$\psi(x) = e^{\frac{-i}{\hbar}\int f(x)dx} \Phi(x) \tag{5}$$

Then $\Phi(x)$ satisfies

$$\frac{d^2\Phi}{dx^2} + \left(\frac{2m(E - V(x))}{\hbar^2}\right)\Phi(x) = 0 \;, \tag{6}$$

which is the standard energy eigenvalue equation for a potential V(x). Thus, the use of the general representation (3) changes neither the energy eigenvalues nor the probability density. It just adds a phase change to the wave function.

Time

As given in reference (2), the symmetrical free particle time operator is:

$$\frac{m}{2}\left(\frac{1}{p}q + q\frac{1}{p}\right). \tag{7}$$

The eigenvalue equation for time is then:

$$\frac{m}{2}\left(\frac{1}{p}q + q\frac{1}{p}\right) = \tau, \tag{8}$$

where $\tau$ is a number. For $\tau \neq 0$, we get

$$p^2 - \left(\frac{m}{\tau}\right)qp + i\left(\frac{\hbar m}{2\tau}\right) = 0. \tag{9}$$

Using (3), this is

$$\frac{d^2\psi}{dx^2} + i\left[\frac{2}{\hbar}f(x) - \frac{m}{\hbar\tau}x\right]\frac{d\psi}{dx} + \left[\frac{i}{\hbar}\frac{df}{dx} + \frac{m}{\tau\hbar^2}xf(x) - \frac{1}{\hbar^2}f^2(x) - \frac{im}{2\tau\hbar}\right]\psi(x) = 0 \tag{10}$$

Let

$$\psi(x) = e^{i\left(\frac{m}{4\hbar\tau}x^2 - \frac{1}{\hbar}\int f(x)dx\right)}\Theta(x). \tag{11}$$

Define

$$\alpha \equiv \frac{m}{\hbar\tau}; \quad y \equiv \sqrt{\alpha}x. \tag{12}$$

Then y is dimensionless and

$$\psi(y) = e^{i\left(\frac{y^2}{4} - \frac{1}{\hbar\sqrt{\alpha}}\int f(y)dy\right)}\Theta(y), \tag{13}$$

where $\Theta(y)$ satisfies

$$\frac{d^2\Theta}{dy^2} + \frac{y^2}{4}\Theta(y) = 0. \tag{14}$$

Just as for energy, the use of the general representation (3) changes neither the time eigenvalues nor the probability density. It only adds a phase change to the wave function.

## The infinite, one-dimensional square well

The solution to (14) is a parabolic cylinder function, but it will be more useful to solve it directly.

The square well runs from 0 to L. Since the walls are infinitely high, any wave function must be 0 at the walls. For $\Theta(0) = 0$, the solution to (14) is:

$$\Theta(y) = a_1 y + \sum_{j=1}^{\infty} a_{4j+1} y^{4j+1} \tag{15}$$

where

$$a_{4j+1} = \frac{(-1)^j a_1}{4^j (4 \cdot 5)(8 \cdot 9)(12 \cdot 13) \cdots (4j)(4j+1)} = \frac{(-1)^j a_1}{4^{2j} \cdot j! \cdot 5 \cdot 9 \cdot 13 \cdots (4j+1)} \tag{16}$$

and $a_1$ is an arbitrary constant. The infinite series for $\Theta(y)$ converges for all y.

## Zeros

The zeros of $\Theta(y)$ are not evenly spaced. [relative error = (value - predicted value)/value.] The nth predicted value is given by:

$$\sqrt{4(n-1)\pi - \pi^{\frac{1}{2}}}, \tag{17}$$

where n is the zero number, except for the first predicted value which is 0 because that is a boundary condition on $\Theta(y)$. Here are the first 60 zeros:

| Zero Number (n) | Zero Position | Difference in Zero Positions | Predicted Zero Positions | Zero Position - Predicted Position | Relative Error |
|---|---|---|---|---|---|
| 1 | 0 | | 0 | 0 | 0 |
| | | 3.3352 | | | |
| 2 | 3.3352 | | 3.335678509 | -0.000478509 | -0.000143472 |
| | | 1.52531 | | | |
| 3 | 4.86051 | | 4.867558087 | -0.007048087 | -0.001450072 |
| | | 1.1536 | | | |
| 4 | 6.01411 | | 6.021585534 | -0.007475534 | -0.001242999 |
| | | 0.96625 | | | |
| 5 | 6.98036 | | 6.98755057 | -0.00719057 | -0.001030114 |
| | | 0.84814 | | | |
| 6 | 7.8285 | | 7.835319622 | -0.006819622 | -0.000871128 |

| | | | | |
|---|---|---|---|---|
| | | 0.76497 | | |
| 7 | 8.59347 | 8.599918848 | -0.006448848 | -0.000750436 |
| | | 0.70229 | | |
| 8 | 9.29576 | 9.301880176 | -0.006120176 | -0.000658384 |
| | | 0.65287 | | |
| 9 | 9.94863 | 9.954463593 | -0.005833593 | -0.000586371 |
| | | 0.61257 | | |
| 10 | 10.5612 | 10.56682147 | -0.005621473 | -0.000532276 |
| | | 0.579 | | |
| 11 | 11.1402 | 11.14558597 | -0.005385971 | -0.000483472 |
| | | 0.5504 | | |
| 12 | 11.6906 | 11.69574526 | -0.005145263 | -0.00044012 |
| | | 0.5256 | | |
| 13 | 12.2162 | 12.22116311 | -0.004963115 | -0.000406273 |
| | | 0.5039 | | |
| 14 | 12.7201 | 12.72490466 | -0.004804655 | -0.000377722 |
| | | 0.4847 | | |
| 15 | 13.2048 | 13.20944999 | -0.004649993 | -0.000352144 |
| | | 0.4675 | | |
| 16 | 13.6723 | 13.67683954 | -0.004539537 | -0.000332024 |
| | | 0.4521 | | |
| 17 | 14.1244 | 14.12877597 | -0.004375967 | -0.000309816 |
| | | 0.438 | | |
| 18 | 14.5624 | 14.56669767 | -0.004297668 | -0.000295121 |
| | | 0.4253 | | |
| 19 | 14.9877 | 14.99183283 | -0.004132829 | -0.000275748 |
| | | 0.4135 | | |
| 20 | 15.4012 | 15.40524009 | -0.004040088 | -0.000262323 |
| | | 0.4027 | | |
| 21 | 15.8039 | 15.8078396 | -0.003939599 | -0.00024928 |
| | | 0.3927 | | |
| 22 | 16.1966 | 16.20043714 | -0.003837136 | -0.00023691 |
| | | 0.3834 | | |
| 23 | 16.58 | 16.58374306 | -0.003743064 | -0.000225758 |
| | | 0.3747 | | |
| 24 | 16.9547 | 16.95838744 | -0.003687442 | -0.000217488 |
| | | 0.3666 | | |
| 25 | 17.3213 | 17.32493219 | -0.003632186 | -0.000209695 |
| | | 0.359 | | |
| 26 | 17.6803 | 17.68388096 | -0.003580962 | -0.00020254 |
| | | 0.3519 | | |
| 27 | 18.0322 | 18.0356873 | -0.003487303 | -0.000193393 |
| | | 0.3451 | | |
| 28 | 18.3773 | 18.38076133 | -0.003461331 | -0.000188348 |
| | | 0.3388 | | |
| 29 | 18.7161 | 18.71947536 | -0.003375359 | -0.000180345 |
| | | 0.3327 | | |
| 30 | 19.0488 | 19.0521686 | -0.003368599 | -0.00017684 |
| | | 0.3271 | | |

| | | | | | |
|---|---|---|---|---|---|
| 31 | 19.3759 | | 19.37915114 | -0.003251141 | -0.000167793 |
| | | 0.3216 | | | |
| 32 | 19.6975 | | 19.70070734 | -0.003207336 | -0.00016283 |
| | | 0.3164 | | | |
| 33 | 20.0139 | | 20.01709869 | -0.003198695 | -0.000159824 |
| | | 0.3115 | | | |
| 34 | 20.3254 | | 20.32856637 | -0.003166373 | -0.000155784 |
| | | 0.3068 | | | |
| 35 | 20.6322 | | 20.63533332 | -0.003133324 | -0.000151866 |
| | | 0.3024 | | | |
| 36 | 20.9346 | | 20.93760617 | -0.003006167 | -0.000143598 |
| | | 0.298 | | | |
| 37 | 21.2326 | | 21.23557681 | -0.002976814 | -0.0001402 |
| | | 0.2939 | | | |
| 38 | 21.5265 | | 21.52942389 | -0.002923895 | -0.000135828 |
| | | 0.2899 | | | |
| 39 | 21.8164 | | 21.81931401 | -0.00291401 | -0.00013357 |
| | | 0.2861 | | | |
| 40 | 22.1025 | | 22.10540283 | -0.002902834 | -0.000131335 |
| | | 0.2825 | | | |
| 41 | 22.385 | | 22.3878361 | -0.002836096 | -0.000126696 |
| | | 0.2789 | | | |
| 42 | 22.6639 | | 22.66675044 | -0.002850444 | -0.00012577 |
| | | 0.2756 | | | |
| 43 | 22.9395 | | 22.94227422 | -0.002774218 | -0.000120936 |
| | | 0.2723 | | | |
| 44 | 23.2118 | | 23.21452814 | -0.002728143 | -0.000117533 |
| | | 0.2691 | | | |
| 45 | 23.4809 | | 23.48362595 | -0.002725945 | -0.000116092 |
| | | 0.2661 | | | |
| 46 | 23.747 | | 23.74967491 | -0.002674906 | -0.000112642 |
| | | 0.2631 | | | |
| 47 | 24.0101 | | 24.01277637 | -0.002676365 | -0.000111468 |
| | | 0.2603 | | | |
| 48 | 24.2704 | | 24.27302617 | -0.002626169 | -0.000108205 |
| | | 0.2575 | | | |
| 49 | 24.5279 | | 24.53051508 | -0.002615078 | -0.000106616 |
| | | 0.2548 | | | |
| 50 | 24.7827 | | 24.78532914 | -0.002629141 | -0.000106088 |
| | | 0.2523 | | | |
| 51 | 25.035 | | 25.03755002 | -0.002550024 | -0.000101858 |
| | | 0.2497 | | | |
| 52 | 25.2847 | | 25.28725532 | -0.002555324 | -0.000101062 |
| | | 0.2473 | | | |
| 53 | 25.532 | | 25.53451884 | -0.002518841 | -9.86543E-05 |
| | | 0.2449 | | | |
| 54 | 25.7769 | | 25.77941084 | -0.002510836 | -9.74064E-05 |
| | | 0.2426 | | | |
| 55 | 26.0195 | | 26.02199826 | -0.002498265 | -9.60151E-05 |

| | | 0.2404 | | | |
|---|---|---|---|---|---|
| 56 | 26.2599 | | 26.26234499 | -0.002444988 | -9.31073E-05 |
| | | 0.2382 | | | |
| 57 | 26.4981 | | 26.50051197 | -0.002411974 | -9.10244E-05 |
| | | 0.236 | | | |
| 58 | 26.7341 | | 26.73655747 | -0.002457473 | -9.19228E-05 |
| | | 0.234 | | | |
| 59 | 26.9681 | | 26.97053719 | -0.002437187 | -9.03729E-05 |
| | | 0.232 | | | |
| 60 | 27.2001 | | 27.20250442 | -0.002404421 | -8.83975E-05 |

The zeros draw steadily closer together and the error in the predictions fall steadily. This is illustrated by the following figures:

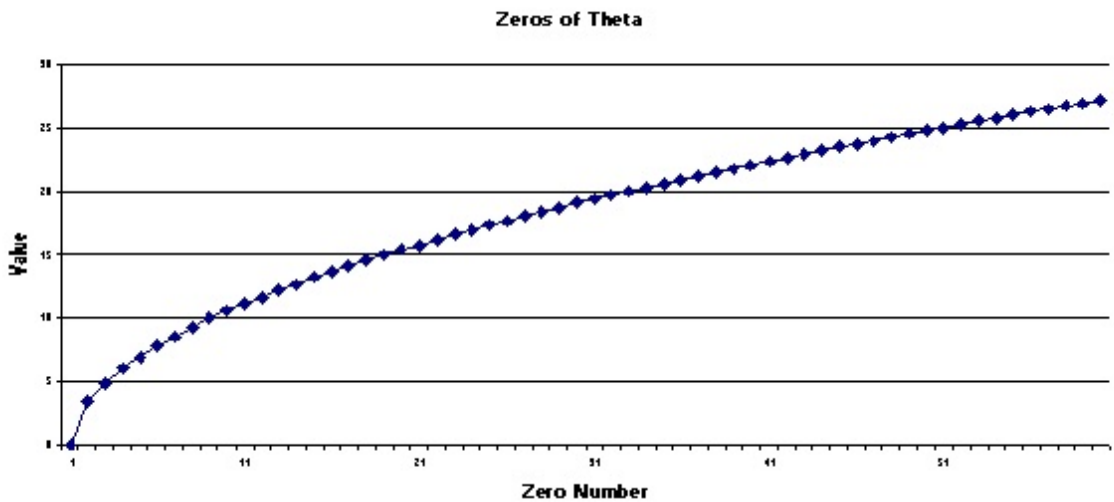

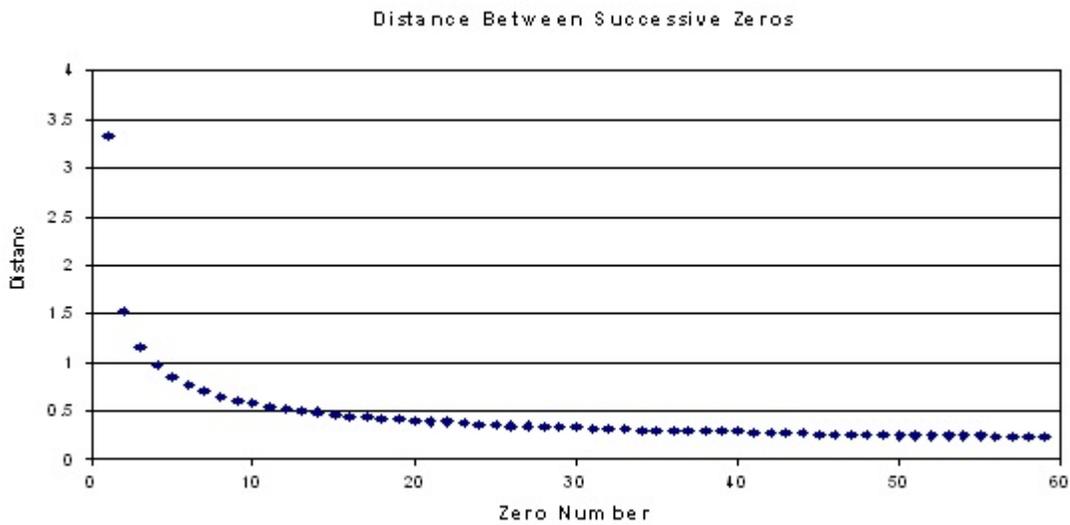

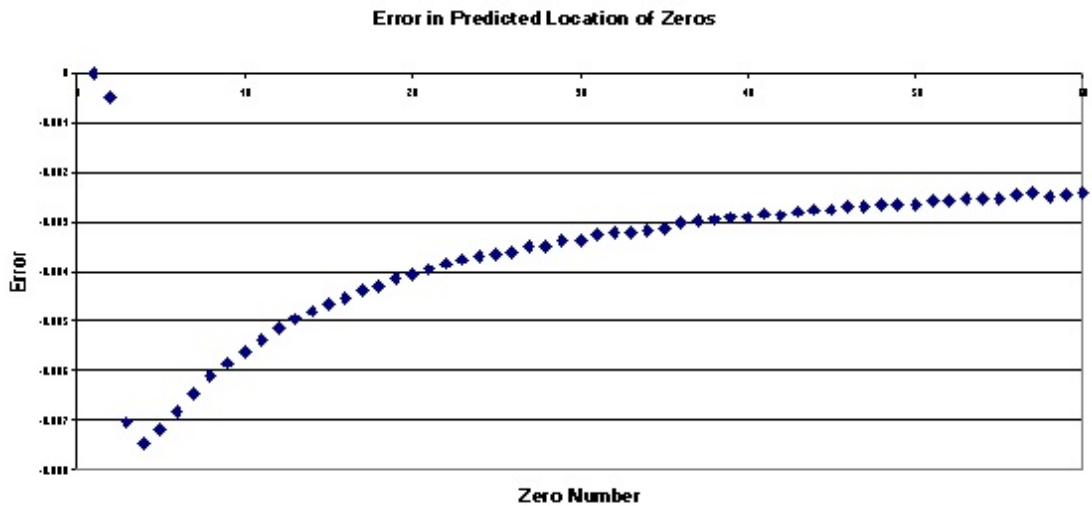

Error in Predicted Location of Zeros

### The Θ Function

The maxima and minima of Θ(y) approach zero from the top and bottom as y goes to infinity. We have already discussed the position of the zeros of Θ(y) which, as n→∞, seem to approach

$$\sqrt{4(n-1)\pi - \pi^{\frac{1}{2}}} \qquad (17)$$

The maxima and minima of Θ have an even simpler pattern as illustrated by the following data. Predicted maximum and minimum values are give by:

$$\Theta(y) = \pm \frac{2}{\sqrt{y}}. \qquad (18)$$

Minima

| y | Value | Predicted | Error |
|---|---|---|---|
| 4.13959 | -0.9791 | -0.983 | 0.003924 |
| 6.5079 | -0.78345 | -0.78399 | 0.000539 |
| 8.21628 | -0.69755 | -0.69774 | 0.00019 |
| 9.62549 | -0.64455 | -0.64464 | 9.4E-05 |
| 10.853 | -0.60704 | -0.60709 | 5.58E-05 |
| 11.9551 | -0.57840 | -0.57843 | 3.54E-05 |
| 12.9638 | -0.55545 | -0.55547 | 2.51E-05 |

Maxima

| y | Value | Predicted | Error |
|---|---|---|---|
| 2.05768 | 1.3356 | 1.394251 | -0.05865 |
| 5.45544 | 0.855098 | 0.856279 | -0.00118 |
| 7.41164 | 0.734335 | 0.734637 | -0.0003 |
| 8.94871 | 0.668445 | 0.668574 | -0.00013 |
| 10.2577 | 0.624391 | 0.624461 | -7E-05 |
| 11.4174 | 0.591854 | 0.591897 | -4.3E-05 |
| 12.4697 | 0.566344 | 0.566372 | -2.8E-05 |

## Time Eigenvalues

The time wave function must be zero at x=L. Thus $\Theta(L)=0$. Let $z_n$ be the dimensionless position of the nth zero of $\Theta$.

Then, from (12), at the wall at x = L:

$$z_n = \sqrt{\alpha_n} L, \tag{19}$$

which gives the eigenvalues of time as

$$\tau_n = \frac{mL^2}{\hbar z_n^2}. \tag{20}$$

For almost all $z_n$, the approximation in (17) is very good and we have, approximately,

$$\tau_n \approx \frac{mL^2}{\hbar \left[ 4(n-1)\pi - \pi^{\frac{1}{\pi}} \right]}. \tag{21}$$

The number of physical significance is the difference in time eigenvalues. The difference between the nth and the (n+k)th eigenvalues is:

$$\tau_n - \tau_{n+k} = \frac{mL^2}{\hbar} \left( \frac{z_{n+k}^2 - z_n^2}{z_n^2 z_{n+k}^2} \right), \tag{22}$$

with the very good approximation

$$\tau_n - \tau_{n+k} = \frac{mL^2}{\hbar} \left[ \frac{4\pi k}{\left(4(n-1)\pi - \pi^{\frac{1}{\pi}}\right)\left(4(n-1+k)\pi - \pi^{\frac{1}{\pi}}\right)} \right]. \tag{23}$$

The Uncertainty Principle

Although the lowest non-zero eigenvalues of energy and time for the infinite square well are not the same as the uncertainties, $\Delta E$ and $\Delta t$, we would expect that the product of the lowest eigenvalues would be on the order of $\hbar$. Thus

$$\left[\frac{\hbar^2 \pi^2}{2mL^2}\right]\left[\frac{mL^2}{\hbar z_2^2}\right] = \frac{\hbar \pi^2}{2z_2^2} = 0.443635188\,\hbar\,. \qquad (24)$$